\documentclass[%
 reprint,
 amsmath,amssymb,
 aps, prl,
 showpacs,
]{revtex4-1}

\usepackage{graphicx}
\usepackage{natbib}
\usepackage{epsfig}
\usepackage{rotate}
\usepackage{float}
\usepackage{amssymb}
\usepackage{amsbsy}
\usepackage{color}
\usepackage{overpic}
\definecolor{seb}{RGB}{0,133,255}
\definecolor{sus}{RGB}{208,32,144}
\usepackage{marvosym}
\usepackage{textcomp}
\usepackage{amsmath}
\usepackage{graphicx}
\usepackage{dcolumn}
\usepackage{bm}

\newcommand\Nu{\nobreak\mbox{$\mathcal N$\hskip-0.95mm$u$}}
\newcommand\Ra{\nobreak\mbox{$\mathcal R$\hskip-0.3mm$a$}}
\newcommand\Rey{\nobreak\mbox{$\mathcal R$\hskip-0.3mm$e$}}

\newcommand\Pran{\nobreak\mbox{$\mathcal P$\hskip-0.3mm$r$}}

\begin{document}

\preprint{APS/123-QED}

\title{Thermal Boundary Layer Equation for Turbulent Rayleigh--B\'enard Convection}%

\author{Olga Shishkina}
 \email{Olga.Shishkina@ds.mpg.de}
\affiliation{Max Planck Institute for Dynamics and Self-Organization,
Am Fassberg 17, D-37077 G\"ottingen, Germany}%
\author{Susanne Horn}
\thanks{present affiliation:
Department of Mathematics, Imperial College London, London SW7 2AZ, UK}
\author{Sebastian Wagner}
\affiliation{Max Planck Institute for Dynamics and Self-Organization, Am Fassberg 17, D-37077 G\"ottingen, Germany}%
\author{Emily S. C. Ching}
 \email{Ching@phy.cuhk.edu.hk}
\affiliation{Department of Physics, The Chinese University of Hong Kong, Shatin, Hong Kong}%
\date{\today}

\begin{abstract}
We report a new thermal boundary layer equation for turbulent
Rayleigh--B\'enard convection for Prandtl number $\Pran>1$ that
takes into account the effect of turbulent fluctuations. These
fluctuations are neglected in existing equations, which are based
on steady-state and laminar assumptions. Using this new equation,
we derive analytically the mean temperature profiles in two
limits: (a) $\Pran \gtrsim 1$ and (b) $\Pran\gg1$. These two
theoretical predictions are in excellent agreement with the
results of our direct numerical simulations for $\Pran=4.38$
(water) and $\Pran=2547.9$ (glycerol) respectively.

\end{abstract}
\pacs{
44.20.+b, 
44.25.+f, 
47.27.ek, 
47.27.te 
}

\maketitle


Turbulent Rayleigh--B\'enard convection (RBC) \cite{Castaing1989, Siggia1994, Ahlers2009, Lohse2010, Chilla2012}, consisting of a fluid confined between two horizontal plates,
heated from below and cooled from above, is a system of great research interest.
It is a paradigm system for studying turbulent thermal convection, which is ubiquitous in nature, occurring in the atmosphere and the mantle of the Earth as well as in stars like our Sun.
Convective heat transfer is also an important problem in engineering and technological applications.
The state of fluid motion in RBC is determined by the Rayleigh number $\Ra = \alpha g \Delta H^3/(\kappa \nu)$ and Prandtl number $\Pran=\nu/\kappa$.
Here $\alpha$ denotes the isobaric thermal expansion coefficient,
$\nu$ the kinematic viscosity and
$\kappa$ the thermal diffusivity of the fluid,
$g$ the acceleration due to gravity,
$\Delta$ the temperature difference between the bottom and top plates,
and $H$ the distance between the plates.

In turbulent RBC, there are viscous boundary layers (BLs) near all
rigid walls and two thermal BLs, one above the bottom plate and
one below the top plate. We denote the thicknesses of the viscous
and thermal BLs by $l$ and $\lambda$ respectively. Both viscous
and thermal BLs play a critical role in the turbulent heat
transfer of the system and in particular $\lambda$ is inversely
proportional to the heat transport. Grossmann and Lohse (GL)
\cite{Grossmann2000}, \cite{Stevens2013} developed a scaling
theory of how the Reynolds number $\Rey$, determined by the mean
large-scale circulation velocity $U_0$
 above the viscous BL, and the dimensionless Nusselt
number $\Nu$, measuring the heat transport, depend on $\Ra$ and
$\Pran$ for moderate $\Ra$. The theory makes explicit use of the
result $l/H \propto\Rey^{-1/2}$ with the proportionality constant
depending only on $\Pran$. This result follows from the
assumptions that the BLs are laminar and their mean profiles,
averaged over time, are described by the
Prandtl--Blasius--Pohlhausen (PBP) theory \cite{Prandtl1905,
Pohlhausen1921, Landau1987} for steady-state forced convection
above an infinite weakly-heated plate. Although the GL theory
gives perfect agreement with the heat transport measurements, the
assumption that the BLs are described by PBP theory is not
fulfilled. Systematic deviations of the mean velocity and
temperature profiles from the PBP predictions have been reported
both in experiments and in direct numerical simulations (DNS)
\cite{Shishkina2009, Shi2012, Scheel2012, Stevens2012,
Kaczorowski2011}. These deviations remain even after a dynamical
rescaling procedure \cite{Zhou2010} that takes into account of the
time variations of $\lambda$ is used, and increase with growing
$\Ra$ and decreasing $\Pran$. An extension of the PBP approach to
the Falkner--Skan--Pohlhausen one \cite{Falkner1931,
Shishkina2013, Shishkina2014}, which accounts for a non-parallel
mean large-scale circulation velocity above the viscous BL
\citep{Wagner2012} and a non-zero pressure gradient within the
BLs, gives better approximations of $l$ and $\lambda$
\cite{Shishkina2014} and is promising for studying mixed
convection \cite{Bailon2012, Shishkina2012} but does not lead to
better predictions of the mean temperature profiles in RBC. For
large $\Pran$, the thermal BL is nested within the viscous one.
Taking the velocity field to be a simple shear flow with constant
shear rate, \citet{Shraiman1990} obtained results for the mean
temperature profile and the relation between the heat flux and
shear rate. Their mean temperature profile coincides with the PBP
prediction for $\Pran \gg 1$. \citet{Ching1997} generalized their
work to study shear flows with position-dependent shear rate, and
obtained mean temperature profiles in terms of two constants that
are functions of $\lambda$, the shear rate and their spatial
derivatives. Good agreement of the derived profile with the actual
ones can be obtained only when these two constants are treated as
free fitting parameters with no solid theoretical support.

The observed deviations between the actual profiles and the
existing predictions from laminar BL models are the effects of
turbulent fluctuations. As $\Ra$ increases, the present
understanding is that the thermal BLs would eventually become
turbulent such that a clear distinction between the BLs and the
bulk of the flow ceases to exist. In this asymptotic state, known
as the ultimate regime, logarithmic mean temperature profiles are
predicted \cite{Grossmann2012} based on the idea of eddy thermal
diffusivity~\cite{Landau1987}. Recently, logarithmic mean
temperature profiles in the turbulent bulk region have also been
reported for  moderate $\Ra$ \cite{Ahlers2014}. 
The separation of the region close to the plate into a viscous
sublayer and a fluctuating logarithmic layer \cite{Grossmann2012,
Ahlers2014} gives a good description of the mean temperature profile
in these two separate subregions but a universal function
predicting correctly the mean temperature profile throughout the
whole region, from the plate to the edge of the bulk region,
remains lacking.

In this Letter, we report a new thermal BL equation for turbulent
RBC that takes into account the effect of the turbulent
fluctuations, which are neglected in the existing BL equations
based on steady-state and laminar assumptions.  Using this
equation, we derive analytically the mean temperature profiles for
$\Pran\gtrsim 1$ and $\Pran\gg1$.
We have performed DNS for $\Pran = 4.38$ (water) and
$\Pran = 2547.9$ (glycerol) with $\Ra$ between $10^7$ and $10^{10}$ in a cylindrical domain of aspect ratio one,
using well-tested finite-volume codes.
The DNS for $\Ra$ up to $10^9$ were conducted using the RBC-version \cite{flowsi2} of the code \cite{flowsi1}.
The simulations for higher $\Ra$  were obtained using our new code {\it goldfish},
which features a versatile operator approach, a high modularity and fully parallel I/O and was 
validated against \cite{flowsi2} for $\Ra = 10^8$.
The computational grids used resolve Kolmogorov and Batchelor scales
in the whole domain \cite{Shishkina2010}. 
Our theoretical predictions are in excellent agreement with our DNS results.

We consider the fluid flow along an infinite horizontal heated plate  and assume that far away from the plate there exists
a constant horizontal mean velocity, the wind, along a certain preferential direction.
We set up the coordinate system such that the $x$-direction is along the wind and $z$-direction is vertical away from the plate.
As the dependence of the mean flow on the other horizontal direction is weak when the plate is large,
we consider a two-dimensional flow that depends on $x$ and $z$ only.
Denote the velocity field by $u(x,z,t)\hat{x} + v(x,z,t) \hat{z}$ and the temperature field by $T(x,z,t)$, where $\hat{x}$ and $\hat{z}$ are
the unit vectors in the corresponding directions and $t$ is the time.
Close to that plate the equation of motion of temperature is governed by
\begin{equation}
\label{tempBL}
\partial_t T + u \partial_x T + v \partial_z T = \kappa \partial_z^2 T,
\end{equation}
where we have used the BL approximation of $|\partial_x^2 T| \ll |\partial_z^2 T|$.
Applying Reynolds decomposition, we can write the velocity and temperature fields as the sums of their long time averages, denoted by $U(x,z)$,
$V(x,z)$ and $\Theta(x,z)$, and their fluctuations defined by
\begin{eqnarray}
\label{51}
u(x,z,t) &=& U(x,z) + u'(x,z,t), \\
\label{521}
 v(x,z,t) &=& V(x,z) + v'(x,z,t), \\
\label{53}
  T(x,z,t) &=& \Theta(x,z) + \theta'(x,z,t).
\end{eqnarray}
Here $U(x,z) \to U_0$ as $z \to \infty$.
Taking a long time average, denoted by $\langle\cdot\rangle_t$, of (\ref{tempBL}), we obtain
\begin{equation}
\label{6} U \partial_x \Theta +
 V \partial_z \Theta +
\partial_x \langle u' \theta'
 \rangle_t + \partial_z \langle v'  \theta'
\rangle_t = \kappa  \partial^2_z \Theta,
\end{equation}
Assuming that $|\partial_x\langle u' \theta'
\rangle_t|\ll|\partial_z \langle v' \theta' \rangle_t|$ and using
the eddy thermal diffusivity $\kappa_{t}=\kappa_{t}(x,z)$, defined
as
\begin{equation}
\label{10} \langle v' \theta' \rangle_{t} \equiv -\kappa_{t}
\partial_z \Theta ,
\end{equation}
one obtains the following BL equation
\begin{equation}
\label{14} U \partial_x \Theta+ (V-\partial_z \kappa_{t})
\partial_z \Theta = (\kappa+\kappa_{t}) \partial_z^2 \Theta.
\end{equation}
We seek a similarity solution of the BL equation (\ref{14}) with
respect to the similarity variable $\xi$, defined by
\begin{equation}
\label{15} \xi=z/\lambda(x),
\end{equation}
where $\lambda(x)$ is the local thickness of the thermal BL.
Let the stream function $\Psi(x,z)$ for the mean velocity be
\begin{equation}
\label{16} \Psi(x,z)=U_0 \lambda(x) \psi(\xi),
\end{equation}
such that $U=\partial_z\Psi$ and  $V=-\partial_x\Psi$, and
$\Theta(x,z)$ be
\begin{equation}
\label{18} \Theta=T_{bot}-(\Delta/2) \theta(\xi) .
\end{equation}
Here $T_{bot}$ is the temperature of the bottom plate.
The boundary conditions for $\psi$ and $\theta$ are
\begin{eqnarray}
\psi(0)=0, \qquad \psi_\xi(0) &=& 0, \qquad \psi_\xi(\infty) = 1, \label{velBC} \\
\theta(0) = 0, \qquad \theta_\xi(0) &=& 1, \qquad
\theta(\infty) = 1. \label{BCs}
\end{eqnarray}
Here the subscript $\xi$ denotes the derivative with respect to $\xi$.
Using (\ref{15})--(\ref{18}) in (\ref{14}) one obtains the following dimensionless BL equation
\begin{eqnarray}
\label{19}
(1+\kappa_{t}/\kappa)\theta_{\xi\xi}+(A+B\psi)\theta_\xi=0,\\
\label{20} A={(\kappa_t)_\xi}/{\kappa},\quad
B={U_0\lambda\lambda_x}/{\kappa}.
\end{eqnarray}
and the subscript $x$ denotes the derivative with respect to $x$.
For the similarity solution to exist, $B$ must be constant, independent of $x$, therefore $\lambda(x) \propto \sqrt{x}$.
In the BL approximation,  $l \propto \lambda$ and thus $l \propto \sqrt{x}$ as in PBP theory and is therefore consistent with the
assumption used in GL scaling theory \cite{Grossmann2000} for moderate $\Ra$.
We write
\begin{equation}
\lambda(x) = f\sqrt{{\nu x}/{U_0}} \label{lambda}
\end{equation}
and thus $B = \Pran f^2/2$, where $f=f(\Pran)$ is some function
of $\Pran$ that is fixed by the requirement $\theta_\xi(0)=1$.

\begin{figure}
\unitlength1.02truecm
\begin{picture}(18.0,5.3)
\put(0.75,0.3){\includegraphics[width=7.905cm]{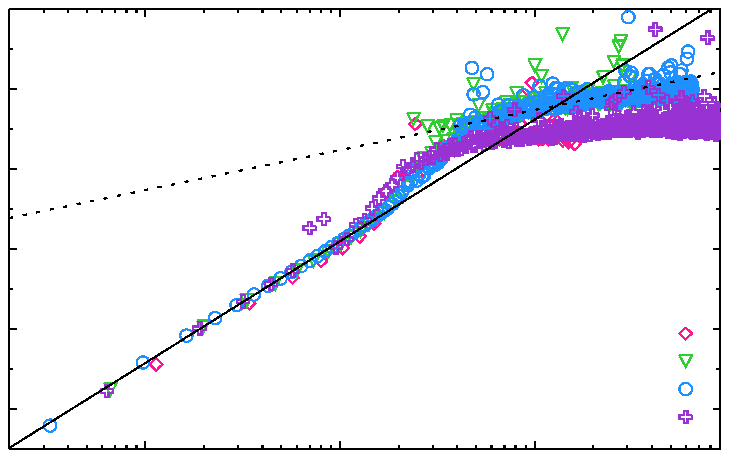}}
\put(0.3,4.1){\small $10^{4}$}
\put(0.3,3.25){\small $10^{2}$}
\put(0,3.5){\rotatebox{90}{$|\kappa_t/\kappa|$}}
\put(0.8,0.8){\rotatebox{31.5}{\small $(1.756\pm0.013)\xi^{3.136\pm0.030}$}}
\put(0.3,2.4){\small $10^{0}$}
\put(0.1,1.55){\small $10^{-2}$}
\put(0.1,0.7){\small $10^{-4}$}
\put(1.95,0){\small $10^{-1}$}
\put(4.15,0){\small $10^{0}$}
\put(6.25,0){\small $10^{1}$}
\put(5.35,-0.2){$\xi$}
\put(6.6,1.5){\small $\Ra=10^7$}
\put(6.6,1.175){\small $\Ra=10^8$}
\put(6.6,0.85){\small $\Ra=10^9$}
\put(6.6,0.52){\small $\Ra=10^{10}$}
\end{picture}
\caption{ 
Normalized eddy thermal diffusivity
$|\kappa_{t}/\kappa|$, 
calculated for $\kappa_{t} = \left(V\Theta-\langle v T\rangle_{t}\right)/\partial_z \Theta$,
and then averaged over horizontal cross-sections, obtained in the DNS 
for $\Pran=4.38$ and $\Ra=10^7$ (diamonds), $10^8$ (triangles),
$10^9$ (circles) and $10^{10}$ (pluses) together with a fit for
$\Ra=10^9$ (solid line). It can be seen that close to the plate,
$\kappa_{t}/\kappa\propto \xi^3$ holds. Dashed line shows the
slope $\propto \xi$ that causes the logarithmical temperature
profiles in the core part of the domain for sufficiently large
$\Ra$.} 
\label{figure:1}
\end{figure}

In the case where fluctuations are ignored, $\langle v'\theta' \rangle_t=0$, $\kappa_t=0$, then (\ref{19}) reduces to the PBP equation.
It was derived in \citet{Shishkina2013} that the PBP equation can be written as
\begin{eqnarray}
\label{2}
\theta_{\xi\xi}+\omega\Gamma^\omega\left(1+\omega^{-1}\right)\xi^{\omega-1}\theta_{\xi}=0
\end{eqnarray}
with $\omega=2$ for $\Pran\ll1$  and $\omega = 3$ for $\Pran\gg1$ and thus all PBP temperature profiles for any $\Pran$ are bounded by
\begin{eqnarray}
\label{1} \theta(\xi) =
\int_0^\xi\exp\left[-\Gamma^\omega\left(1+\omega^{-1}\right)
\chi^\omega\right]\,{\rm d}\chi,
\end{eqnarray}
with $2 \le \omega \le 3$ where $\Gamma$ is the gamma function.
To take into account the fluctuations, we need to know $\kappa_t(\xi)$.
A common approach for fully turbulent BLs is $\kappa_t \propto \xi$ \cite{Landau1987} consequently leading
to logarithmic temperature profiles.
For moderate $\Ra$, such log-profiles are also found but only in the turbulent bulk, which is at a relatively large distance from the plate.
In the vicinity of the plate $\kappa_t$ behaves rather as $\kappa_t\propto \xi^3$ (see Fig.~\ref{figure:1}).
Indeed, from the continuity equation for the fluctuating velocities,
$\partial_x{u'}+
\partial_z{v'}=0$, and that $\theta'$, $ v'$ and $\partial_x u'$ vanishing at the plate, we obtain for $z=0$:
\begin{equation}
\label{22} \langle v' \theta' \rangle_{t}=
\partial_z\langle v' \theta' \rangle_{t}=
\partial^2_z\langle v' \theta' \rangle_{t}=0.
\end{equation}
Using the relation (\ref{10}) and (\ref{15}), (\ref{22}) implies
\begin{equation}
\label{23}
\kappa_{t}(0)=(\kappa_{t})_\xi(0)=(\kappa_{t})_{\xi\xi}(0)=0.
\end{equation}
Therefore the following approximation
\begin{equation}
\label{24}
\kappa_{t}/\kappa\approx a^3\xi^3
\end{equation}
holds for small $\xi$ with some dimensionless constant $a$.
Substituting (\ref{20}) and (\ref{24}) into (\ref{19}), one obtains the following BL equation for the temperature:
\begin{equation}
\label{temp}
(1+a^3\xi^3)\theta_{\xi\xi}+(3a^3\xi^2+B\psi)\theta_\xi=0.
\end{equation}

\begin{figure}
\unitlength1.02truecm
\begin{picture}(18.0,5.3)
\put(0.78,0.3){\includegraphics[width=7.854cm]{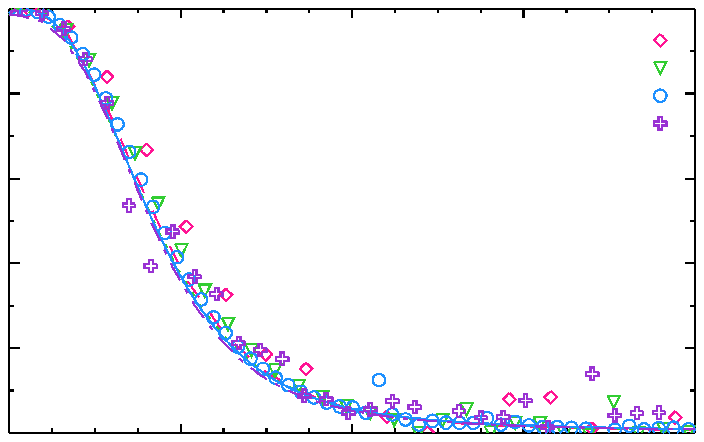}}
\put(0,3.0){\rotatebox{90}{$\kappa/(\kappa_t+\kappa)$}}
\put(0.35,4.9){\small $1.0$}
\put(0.35,3.97){\small $0.8$}
\put(0.35,3.04){\small $0.6$}
\put(0.35,2.11){\small $0.4$}
\put(0.35,1.18){\small $0.2$}
\put(0.35,0.25){\small $0.0$}
\put(0.75,0){\small $0$}
\put(2.65,0){\small $1$}
\put(4.55,0){\small $2$}
\put(6.45,0){\small $3$}
\put(8.35,0){\small $4$}
\put(7.35,-0.2){$\xi$}
\put(6.55,4.6){\small $\Ra=10^7$}
\put(6.55,4.28){\small $\Ra=10^8$}
\put(6.55,3.96){\small $\Ra=10^9$}
\put(6.55,3.64){\small $\Ra=10^{10}$}
\end{picture}
\caption{ 
$\kappa/(\kappa+\kappa_{t})$ (symbols)
[$\kappa_t$ is calculated as in Fig.~\ref{figure:1}],
and $\theta_\xi$ (lines), averaged in time and over horizontal
cross-sections, as functions of the dimensionless vertical
coordinate $\xi$ [see definition in (\ref{15})] obtained in
the DNS for $\Pran=4.38$ 
and $\Ra=10^7$ (diamonds, dashed line), $10^8$ (triangles, dotted
line), $10^9$ (circles, solid line), $10^{10}$ (pluses, dot-dashed
line).} 
\label{figure:2}
\end{figure}

For large $\Pran$, the thermal BL is nested within the viscous BL such that $\lambda < l$ and we can approximate $U \propto z$ within the thermal BL.
Together with (\ref{15})--(\ref{BCs}), one obtains
\begin{equation}
\label{21}
\psi\approx b\xi^2,\quad b=0.5\psi_{\xi\xi}(0).
\end{equation}
(\ref{21}) and (\ref{temp}) lead to the following new thermal BL equation for large $\Pran>1$:
\begin{equation}
\label{25} (1+a^3\xi^3)\theta_{\xi\xi}+(3a^3+bB)\xi^2\theta_\xi=0.
\end{equation}
The solution of (\ref{25}) is
\begin{equation}
\label{26} \theta(\xi)=\int_0^\xi (1+a^3\eta^3)^{-c}d\eta
\end{equation}
with $c=\frac{bB}{3a^3}+1$.
Note that the constants $a$ and $c$ are related by the requirement $\theta(\infty)=1$, which gives
\begin{eqnarray}
\label{a}
a=\frac{\Gamma\left({1}/{3}\right)\Gamma\left(c-{1}/{3}\right)}{3\Gamma\left(c\right)}.
\end{eqnarray}
The order of magnitude of $a$ can be estimated as follows.
Averaging (\ref{6}) in the $x$-direction, denoted by $\langle \cdot \rangle_x$, and integrating it in the vertical
direction from 0 to $z$, and using (\ref{10}) and the definition of the Nusselt number
\begin{equation}
\label{8} \Nu\equiv (\langle vT \rangle_{tx}-\kappa\partial_z
\langle {T} \rangle_{tx})/(\kappa\Delta/H)  ,
\end{equation}
one obtains
\begin{equation}
\label{11} \langle {V} {\Theta}\rangle_{x} =
\langle(\kappa+\kappa_t)
\partial_z  \Theta \rangle_{x} +\Nu\,\kappa\Delta/H.
\end{equation}
Close to the plate, the order of magnitude of the left-hand side of (\ref{11}) is much smaller than $\Nu\,\kappa\Delta/H$, hence,
in this region the following approximation holds:
\begin{equation}
\label{12}
\langle(\kappa+\kappa_t) \partial_z  \Theta +\kappa\Delta/(2\lambda) \rangle_{x}\approx0,
\end{equation}
where we have used the definition \begin{equation}
\lambda(x) \equiv  -\partial_z \Theta \big|_{z=0}/(\Delta/2)
\label{lambdaDef}\end{equation} to write
$\Nu = \langle H/(2\lambda) \rangle_x$.
Approximate that (\ref{12}) holds locally, without the averaging
in $x$, in the region far away from the two vertical walls, we get
\begin{equation}
\label{27} {\kappa}/({\kappa+\kappa_t})\sim
-({2\lambda}/{\Delta})\partial_z \Theta =\theta_\xi
\end{equation}
(see Fig.~\ref{figure:2}). Our DNS show that at the edge ($\xi=1$)
of the thermal BL, $0.36<\theta_\xi<0.65$ holds for all $\Pran$
studied. From (\ref{24}) and (\ref{27}) for $\xi=1$ one obtains
that $0.52<a^3<1.76$, with $a \sim 1.2$ for $\Pran \gtrsim 1$ and
$a \sim 0.8$ for $\Pran \gg 1$. Thus we have $c \sim 1$ for $\Pran
\sim 1$ and $c \sim 2$ for $\Pran \gg 1$.

The analytical solution (\ref{26}) of the BL equation (\ref{25})
that satisfies (\ref{BCs}) for $c=1$ reads
\begin{eqnarray}
\label{28} \theta&=&\frac{\sqrt{3}}{4\pi}
\log\frac{(1+a\xi)^3}{1+(a\xi)^3}+ \frac{3}{2\pi}\arctan\left(
\frac{4\pi}{9}\xi-\frac{1}{\sqrt{3}}
\right)+\frac{1}{4},\nonumber\\
&&a={2\pi}/({3\sqrt{3}})\approx1.2,
\end{eqnarray}
while that for  $c=2$  is
\begin{eqnarray}
\label{29} \theta&=&\frac{\sqrt{3}}{4\pi}
\log\frac{(1+a\xi)^3}{1+(a\xi)^3}+ \frac{3}{2\pi}\arctan\left(
\frac{8\pi}{27}\xi-\frac{1}{\sqrt{3}}
\right)\\
&+&
\frac{\xi}{3(1+(a\xi)^3)}+\frac{1}{4},\qquad
a={4\pi}/({9\sqrt{3}})\approx0.8. \nonumber
\end{eqnarray}
Thus, all temperature profiles for $\Pran>1$ lie between
(\ref{28}) ($\Pran\gtrsim1$) and (\ref{29}) ($\Pran\gg1$).

\begin{figure}
\unitlength1.0truecm
\begin{picture}(18.0,5.6)
\put(0.43,0.3){\includegraphics[width=8.2cm]{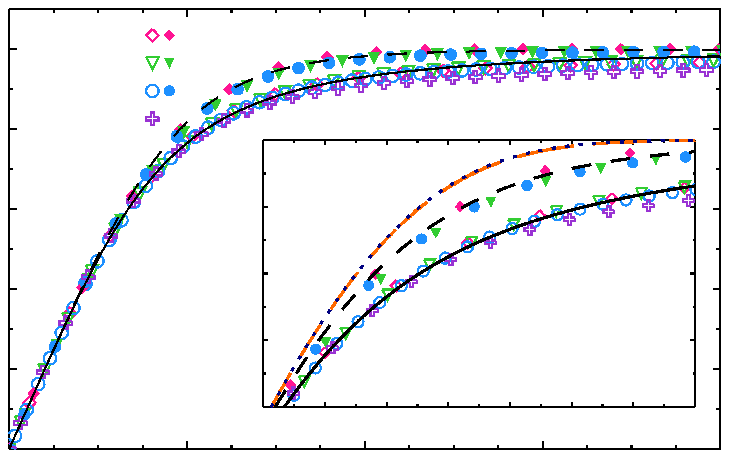}}
\put(-0.1,4.3){$\theta$}
\put(0,4.8){\small $1.0$}
\put(0,3.89){\small $0.8$}
\put(0,2.98){\small $0.6$}
\put(0,2.07){\small $0.4$}
\put(0,1.16){\small $0.2$}
\put(0,0.25){\small $0.0$}
\put(0.4,0){\small $0$}
\put(2.425,0){\small $1$}
\put(4.45,0){\small $2$}
\put(6.475,0){\small $3$}
\put(8.5,0){\small $4$}
\put(7.5,-0.2){$\xi$}
\put(0.65,5.0){\small $\Ra=10^7$}
\put(0.65,4.65){\small $\Ra=10^8$}
\put(0.65,4.3){\small $\Ra=10^9$}
\put(0.55,3.95){\small $\Ra=10^{10}$}
\put(3.2,0.5){\small $0.6$}
\put(3.9,0.5){\small $0.8$}
\put(4.6,0.5){\small $1.0$}
\put(5.3,0.5){\small $1.2$}
\put(6.0,0.5){\small $1.4$}
\put(6.7,0.5){\small $1.6$}
\put(7.4,0.5){\small $1.8$}
\put(8.1,0.5){\small $2.0$}
\put(2.9,3.75){\small $1.0$}
\put(2.9,3){\small $0.9$}
\put(2.9,2.25){\small $0.8$}
\put(2.9,1.5){\small $0.7$}
\put(2.9,0.75){\small $0.6$}
\end{picture}
\caption{
Temperature profiles, averaged in time and over horizontal 
cross-sections, obtained in the DNS 
for $\Pran=4.38$ (open symbols) and $\Pran=2547.9$ (filled
symbols) and $\Ra=10^7$ (diamonds), $10^8$ (triangles), $10^9$
(circles) and $10^{10}$ (pluses). Excellent agreement with the
predictions for $\Pran\gtrsim1$ (\ref{28}) (solid line) and
$\Pran\gg1$ (\ref{29}) (dashed line) is demonstrated. An expanded
view with the PBP prediction (\ref{1}) for $\Pran=2547.9$ (dotted
line) and $\Pran=4.38$ (long-dashed orange line, which almost
coincides with the dotted line) for comparison is shown in the
inset.} 
\label{figure:3}
\end{figure}

Next we compare our predictions (\ref{28}),
(\ref{29}) with the DNS results.
At each of the two $\Pran$ (4.38 and 2547.9) studied, the mean
temperature profiles almost collapse for the different $\Ra$
($10^7$ to $10^{9}$). 
Generally the profiles depend
very weakly on $\Ra$ (see Fig.~\ref{figure:3}). 
For $\Ra$ from $10^7$ to $10^{9}$, the DNS
profiles for $\Pran=4.38$ are in perfect agreement with the
predicted profile (\ref{28}) for $\Pran \gtrsim 1$,  while the DNS
profiles for $\Pran=2547.9$ are in perfect agreement with the
predicted profile (\ref{29}).
On the other hand, the PBP
predictions for $\Pran=2547.9$ and $\Pran=4.38$ almost coincide
with the PBP prediction for $\Pran \gg 1$ and lie well above the
corresponding DNS profiles. 

In summary, we have derived a new thermal BL equation for
turbulent RBC for $\Pran>1$ using the idea of an eddy thermal
diffusivity, which close to the plate 
is shown to depend on the cubic power of the
distance from the plate. We have solved the equation to obtain two
analytical mean temperature profiles for $\Pran \gtrsim 1$ and
$\Pran \gg 1$ respectively, and demonstrated that they are in
excellent agreement with the DNS profiles. 
The general dependence of the coefficient $a$ and thus the temperature
profile (\ref{26}) on $\Pran$, $\Ra$, and the geometrical
characteristics of the convection cell will be explored in future
studies.

\acknowledgments
The authors OS, SH, and SW acknowledge financial support of the Deutsche Forschungsgemeinschaft (DFG) under the grants SH405/3 and SFB 963/1, project A6, and the
Heisenberg fellowship SH405/4 while the work of ESCC is supported by the Hong Kong Research Grants Council under Grant No. CUHK-400311.
The authors thank also the Leibniz--Rechenzentrum (LRZ) in Garching for providing computational resources on SuperMUC.
\bibliography{sebref}

\begin{thebibliography}{29}%
\makeatletter
\providecommand \@ifxundefined [1]{%
 \@ifx{#1\undefined}
}%
\providecommand \@ifnum [1]{%
 \ifnum #1\expandafter \@firstoftwo
 \else \expandafter \@secondoftwo
 \fi
}%
\providecommand \@ifx [1]{%
 \ifx #1\expandafter \@firstoftwo
 \else \expandafter \@secondoftwo
 \fi
}%
\providecommand \natexlab [1]{#1}%
\providecommand \enquote  [1]{``#1''}%
\providecommand \bibnamefont  [1]{#1}%
\providecommand \bibfnamefont [1]{#1}%
\providecommand \citenamefont [1]{#1}%
\providecommand \href@noop [0]{\@secondoftwo}%
\providecommand \href [0]{\begingroup \@sanitize@url \@href}%
\providecommand \@href[1]{\@@startlink{#1}\@@href}%
\providecommand \@@href[1]{\endgroup#1\@@endlink}%
\providecommand \@sanitize@url [0]{\catcode `\\12\catcode `\$12\catcode
  `\&12\catcode `\#12\catcode `\^12\catcode `\_12\catcode `\%12\relax}%
\providecommand \@@startlink[1]{}%
\providecommand \@@endlink[0]{}%
\providecommand \url  [0]{\begingroup\@sanitize@url \@url }%
\providecommand \@url [1]{\endgroup\@href {#1}{\urlprefix }}%
\providecommand \urlprefix  [0]{URL }%
\providecommand \Eprint [0]{\href }%
\providecommand \doibase [0]{http://dx.doi.org/}%
\providecommand \selectlanguage [0]{\@gobble}%
\providecommand \bibinfo  [0]{\@secondoftwo}%
\providecommand \bibfield  [0]{\@secondoftwo}%
\providecommand \translation [1]{[#1]}%
\providecommand \BibitemOpen [0]{}%
\providecommand \bibitemStop [0]{}%
\providecommand \bibitemNoStop [0]{.\EOS\space}%
\providecommand \EOS [0]{\spacefactor3000\relax}%
\providecommand \BibitemShut  [1]{\csname bibitem#1\endcsname}%
\let\auto@bib@innerbib\@empty
\bibitem [{\citenamefont {Castaing}\ \emph {et~al.}(1989)\citenamefont
  {Castaing}, \citenamefont {Gunaratne}, \citenamefont {Heslot}, \citenamefont
  {Kadanoff}, \citenamefont {Libchaber}, \citenamefont {Thomae}, \citenamefont
  {Wu}, \citenamefont {Zaleski},\ and\ \citenamefont {Zanetti}}]{Castaing1989}%
  \BibitemOpen
  \bibfield  {author} {\bibinfo {author} {\bibfnamefont {B.}~\bibnamefont
  {Castaing}}, \bibinfo {author} {\bibfnamefont {G.}~\bibnamefont {Gunaratne}},
  \bibinfo {author} {\bibfnamefont {F.}~\bibnamefont {Heslot}}, \bibinfo
  {author} {\bibfnamefont {L.}~\bibnamefont {Kadanoff}}, \bibinfo {author}
  {\bibfnamefont {A.}~\bibnamefont {Libchaber}}, \bibinfo {author}
  {\bibfnamefont {S.}~\bibnamefont {Thomae}}, \bibinfo {author} {\bibfnamefont
  {X.-Z.}\ \bibnamefont {Wu}}, \bibinfo {author} {\bibfnamefont
  {S.}~\bibnamefont {Zaleski}}, \ and\ \bibinfo {author} {\bibfnamefont
  {G.}~\bibnamefont {Zanetti}},\ }\href@noop {} {\bibfield  {journal} {\bibinfo
   {journal} {J. Fluid Mech.}\ }\textbf {\bibinfo {volume} {204}},\ \bibinfo
  {pages} {1} (\bibinfo {year} {1989})}\BibitemShut {NoStop}%
\bibitem [{\citenamefont {Siggia}(1994)}]{Siggia1994}%
  \BibitemOpen
  \bibfield  {author} {\bibinfo {author} {\bibfnamefont {E.}~\bibnamefont
  {Siggia}},\ }\href@noop {} {\bibfield  {journal} {\bibinfo  {journal} {Annu.
  Rev. Fluid Mech.}\ }\textbf {\bibinfo {volume} {26}},\ \bibinfo {pages} {137}
  (\bibinfo {year} {1994})}\BibitemShut {NoStop}%
\bibitem [{\citenamefont {Ahlers}\ \emph {et~al.}(2009)\citenamefont {Ahlers},
  \citenamefont {Grossmann},\ and\ \citenamefont {Lohse}}]{Ahlers2009}%
  \BibitemOpen
  \bibfield  {author} {\bibinfo {author} {\bibfnamefont {G.}~\bibnamefont
  {Ahlers}}, \bibinfo {author} {\bibfnamefont {S.}~\bibnamefont {Grossmann}}, \
  and\ \bibinfo {author} {\bibfnamefont {D.}~\bibnamefont {Lohse}},\
  }\href@noop {} {\bibfield  {journal} {\bibinfo  {journal} {Rev. Mod. Phys.}\
  }\textbf {\bibinfo {volume} {81}},\ \bibinfo {pages} {503} (\bibinfo {year}
  {2009})}\BibitemShut {NoStop}%
\bibitem [{\citenamefont {Lohse}\ and\ \citenamefont {Xia}(2010)}]{Lohse2010}%
  \BibitemOpen
  \bibfield  {author} {\bibinfo {author} {\bibfnamefont {D.}~\bibnamefont
  {Lohse}}\ and\ \bibinfo {author} {\bibfnamefont {K.-Q.}\ \bibnamefont
  {Xia}},\ }\href@noop {} {\bibfield  {journal} {\bibinfo  {journal} {Annu.
  Rev. Fluid Mech.}\ }\textbf {\bibinfo {volume} {42}},\ \bibinfo {pages} {335}
  (\bibinfo {year} {2010})}\BibitemShut {NoStop}%
\bibitem [{\citenamefont {Chill\`{a}}\ and\ \citenamefont
  {Schumacher}(2012)}]{Chilla2012}%
  \BibitemOpen
  \bibfield  {author} {\bibinfo {author} {\bibfnamefont {F.}~\bibnamefont
  {Chill\`{a}}}\ and\ \bibinfo {author} {\bibfnamefont {J.}~\bibnamefont
  {Schumacher}},\ }\href@noop {} {\bibfield  {journal} {\bibinfo  {journal}
  {Eur. Phys. J. E}\ }\textbf {\bibinfo {volume} {35}},\ \bibinfo {pages} {58}
  (\bibinfo {year} {2012})}\BibitemShut {NoStop}%
\bibitem [{\citenamefont {Grossmann}\ and\ \citenamefont
  {Lohse}(2000)}]{Grossmann2000}%
  \BibitemOpen
  \bibfield  {author} {\bibinfo {author} {\bibfnamefont {S.}~\bibnamefont
  {Grossmann}}\ and\ \bibinfo {author} {\bibfnamefont {D.}~\bibnamefont
  {Lohse}},\ }\href@noop {} {\bibfield  {journal} {\bibinfo  {journal} {J.
  Fluid Mech.}\ }\textbf {\bibinfo {volume} {407}},\ \bibinfo {pages} {27}
  (\bibinfo {year} {2000})}\BibitemShut {NoStop}%
\bibitem [{\citenamefont {Stevens}\ \emph {et~al.}(2013)\citenamefont
  {Stevens}, \citenamefont {van~der Poel}, \citenamefont {Grossmann},\ and\
  \citenamefont {Lohse}}]{Stevens2013}%
  \BibitemOpen
  \bibfield  {author} {\bibinfo {author} {\bibfnamefont {R.~J. A.~M.}\
  \bibnamefont {Stevens}}, \bibinfo {author} {\bibfnamefont {E.~P.}\
  \bibnamefont {van~der Poel}}, \bibinfo {author} {\bibfnamefont
  {S.}~\bibnamefont {Grossmann}}, \ and\ \bibinfo {author} {\bibfnamefont
  {D.}~\bibnamefont {Lohse}},\ }\href@noop {} {\bibfield  {journal} {\bibinfo
  {journal} {J. Fluid Mech.}\ }\textbf {\bibinfo {volume} {730}},\ \bibinfo
  {pages} {295} (\bibinfo {year} {2013})}\BibitemShut {NoStop}%
\bibitem [{\citenamefont {Prandtl}(1905)}]{Prandtl1905}%
  \BibitemOpen
  \bibfield  {author} {\bibinfo {author} {\bibfnamefont {L.}~\bibnamefont
  {Prandtl}},\ }in\ \href@noop {} {\emph {\bibinfo {booktitle} {Verhandlungen
  des III. Int. Math. Kongr., Heidelberg, 1904}}}\ (\bibinfo  {publisher}
  {Teubner},\ \bibinfo {address} {Leipzig},\ \bibinfo {year} {1905})\ pp.\
  \bibinfo {pages} {484--491}\BibitemShut {NoStop}%
\bibitem [{\citenamefont {Pohlhausen}(1921)}]{Pohlhausen1921}%
  \BibitemOpen
  \bibfield  {author} {\bibinfo {author} {\bibfnamefont {E.}~\bibnamefont
  {Pohlhausen}},\ }\href@noop {} {\bibfield  {journal} {\bibinfo  {journal} {Z.
  Angew. Math. Mech.}\ }\textbf {\bibinfo {volume} {1}},\ \bibinfo {pages}
  {115} (\bibinfo {year} {1921})}\BibitemShut {NoStop}%
\bibitem [{\citenamefont {Landau}\ and\ \citenamefont
  {Lifshitz}(1987)}]{Landau1987}%
  \BibitemOpen
  \bibfield  {author} {\bibinfo {author} {\bibfnamefont {L.~D.}\ \bibnamefont
  {Landau}}\ and\ \bibinfo {author} {\bibfnamefont {E.~M.}\ \bibnamefont
  {Lifshitz}},\ }\href@noop {} {\emph {\bibinfo {title} {Fluid Mechanics}}},\
  \bibinfo {edition} {2nd}\ ed.,\ \bibinfo {series} {Course of theoretical
  physics}, Vol.~\bibinfo {volume} {6}\ (\bibinfo {year} {1987})\BibitemShut
  {NoStop}%
\bibitem [{\citenamefont {Shishkina}\ and\ \citenamefont
  {Thess}(2009)}]{Shishkina2009}%
  \BibitemOpen
  \bibfield  {author} {\bibinfo {author} {\bibfnamefont {O.}~\bibnamefont
  {Shishkina}}\ and\ \bibinfo {author} {\bibfnamefont {A.}~\bibnamefont
  {Thess}},\ }\href@noop {} {\bibfield  {journal} {\bibinfo  {journal} {J.
  Fluid Mech.}\ }\textbf {\bibinfo {volume} {663}},\ \bibinfo {pages} {449}
  (\bibinfo {year} {2009})}\BibitemShut {NoStop}%
\bibitem [{\citenamefont {Shi}\ \emph {et~al.}(2012)\citenamefont {Shi},
  \citenamefont {Emran},\ and\ \citenamefont {Schumacher}}]{Shi2012}%
  \BibitemOpen
  \bibfield  {author} {\bibinfo {author} {\bibfnamefont {N.}~\bibnamefont
  {Shi}}, \bibinfo {author} {\bibfnamefont {M.~S.}\ \bibnamefont {Emran}}, \
  and\ \bibinfo {author} {\bibfnamefont {J.}~\bibnamefont {Schumacher}},\
  }\href@noop {} {\bibfield  {journal} {\bibinfo  {journal} {J. Fluid Mech.}\
  }\textbf {\bibinfo {volume} {706}},\ \bibinfo {pages} {5} (\bibinfo {year}
  {2012})}\BibitemShut {NoStop}%
\bibitem [{\citenamefont {Scheel}\ \emph {et~al.}(2012)\citenamefont {Scheel},
  \citenamefont {Kim},\ and\ \citenamefont {White}}]{Scheel2012}%
  \BibitemOpen
  \bibfield  {author} {\bibinfo {author} {\bibfnamefont {J.~D.}\ \bibnamefont
  {Scheel}}, \bibinfo {author} {\bibfnamefont {E.}~\bibnamefont {Kim}}, \ and\
  \bibinfo {author} {\bibfnamefont {K.~R.}\ \bibnamefont {White}},\ }\href@noop
  {} {\bibfield  {journal} {\bibinfo  {journal} {J. Fluid Mech.}\ }\textbf
  {\bibinfo {volume} {711}},\ \bibinfo {pages} {281} (\bibinfo {year}
  {2012})}\BibitemShut {NoStop}%
\bibitem [{\citenamefont {Stevens}\ \emph {et~al.}(2012)\citenamefont
  {Stevens}, \citenamefont {Zhou}, \citenamefont {Grossmann}, \citenamefont
  {Verzicco}, \citenamefont {Xia},\ and\ \citenamefont {Lohse}}]{Stevens2012}%
  \BibitemOpen
  \bibfield  {author} {\bibinfo {author} {\bibfnamefont {R.~J. A.~M.}\
  \bibnamefont {Stevens}}, \bibinfo {author} {\bibfnamefont {Q.}~\bibnamefont
  {Zhou}}, \bibinfo {author} {\bibfnamefont {S.}~\bibnamefont {Grossmann}},
  \bibinfo {author} {\bibfnamefont {R.}~\bibnamefont {Verzicco}}, \bibinfo
  {author} {\bibfnamefont {K.-Q.}\ \bibnamefont {Xia}}, \ and\ \bibinfo
  {author} {\bibfnamefont {D.}~\bibnamefont {Lohse}},\ }\href@noop {}
  {\bibfield  {journal} {\bibinfo  {journal} {Phys. Rev. E}\ }\textbf {\bibinfo
  {volume} {85}},\ \bibinfo {pages} {027301} (\bibinfo {year}
  {2012})}\BibitemShut {NoStop}%
\bibitem [{\citenamefont {Kaczorowski}\ \emph {et~al.}(2011)\citenamefont
  {Kaczorowski}, \citenamefont {Shishkina}, \citenamefont {Shishkin},
  \citenamefont {Wagner},\ and\ \citenamefont {Xia}}]{Kaczorowski2011}%
  \BibitemOpen
  \bibfield  {author} {\bibinfo {author} {\bibfnamefont {M.}~\bibnamefont
  {Kaczorowski}}, \bibinfo {author} {\bibfnamefont {O.}~\bibnamefont
  {Shishkina}}, \bibinfo {author} {\bibfnamefont {A.}~\bibnamefont {Shishkin}},
  \bibinfo {author} {\bibfnamefont {C.}~\bibnamefont {Wagner}}, \ and\ \bibinfo
  {author} {\bibfnamefont {K.-Q.}\ \bibnamefont {Xia}},\ }in\ \href@noop {}
  {\emph {\bibinfo {booktitle} {Direct and Large-Eddy Simulation VIII}}},\
  \bibinfo {editor} {edited by\ \bibinfo {editor} {\bibnamefont {{H. Kuerten
  et. al.}}}}\ (\bibinfo  {publisher} {Springer},\ \bibinfo {year} {2011})\
  pp.\ \bibinfo {pages} {383--388}\BibitemShut {NoStop}%
\bibitem [{\citenamefont {Zhou}\ and\ \citenamefont {Xia}(2010)}]{Zhou2010}%
  \BibitemOpen
  \bibfield  {author} {\bibinfo {author} {\bibfnamefont {Q.}~\bibnamefont
  {Zhou}}\ and\ \bibinfo {author} {\bibfnamefont {K.-Q.}\ \bibnamefont {Xia}},\
  }\href@noop {} {\bibfield  {journal} {\bibinfo  {journal} {Phys. Rev. Lett.}\
  }\textbf {\bibinfo {volume} {104}},\ \bibinfo {pages} {104301} (\bibinfo
  {year} {2010})}\BibitemShut {NoStop}%
\bibitem [{\citenamefont {Falkner}\ and\ \citenamefont
  {Skan}(1931)}]{Falkner1931}%
  \BibitemOpen
  \bibfield  {author} {\bibinfo {author} {\bibfnamefont {V.~M.}\ \bibnamefont
  {Falkner}}\ and\ \bibinfo {author} {\bibfnamefont {S.~W.}\ \bibnamefont
  {Skan}},\ }\href@noop {} {\bibfield  {journal} {\bibinfo  {journal} {Phil.
  Mag.}\ }\textbf {\bibinfo {volume} {12}},\ \bibinfo {pages} {865} (\bibinfo
  {year} {1931})}\BibitemShut {NoStop}%
\bibitem [{\citenamefont {Shishkina}\ \emph {et~al.}(2013)\citenamefont
  {Shishkina}, \citenamefont {Horn},\ and\ \citenamefont
  {Wagner}}]{Shishkina2013}%
  \BibitemOpen
  \bibfield  {author} {\bibinfo {author} {\bibfnamefont {O.}~\bibnamefont
  {Shishkina}}, \bibinfo {author} {\bibfnamefont {S.}~\bibnamefont {Horn}}, \
  and\ \bibinfo {author} {\bibfnamefont {S.}~\bibnamefont {Wagner}},\
  }\href@noop {} {\bibfield  {journal} {\bibinfo  {journal} {J. Fluid Mech.}\
  }\textbf {\bibinfo {volume} {730}},\ \bibinfo {pages} {442} (\bibinfo {year}
  {2013})}\BibitemShut {NoStop}%
\bibitem [{\citenamefont {Shishkina}\ \emph {et~al.}(2014)\citenamefont
  {Shishkina}, \citenamefont {Wagner},\ and\ \citenamefont
  {Horn}}]{Shishkina2014}%
  \BibitemOpen
  \bibfield  {author} {\bibinfo {author} {\bibfnamefont {O.}~\bibnamefont
  {Shishkina}}, \bibinfo {author} {\bibfnamefont {S.}~\bibnamefont {Wagner}}, \
  and\ \bibinfo {author} {\bibfnamefont {S.}~\bibnamefont {Horn}},\ }\href
  {\doibase 10.1103/PhysRevE.89.033014} {\bibfield  {journal} {\bibinfo
  {journal} {Phys. Rev. E}\ }\textbf {\bibinfo {volume} {89}},\ \bibinfo
  {pages} {033014} (\bibinfo {year} {2014})}\BibitemShut {NoStop}%
\bibitem [{\citenamefont {Wagner}\ \emph {et~al.}(2012)\citenamefont {Wagner},
  \citenamefont {Shishkina},\ and\ \citenamefont {Wagner}}]{Wagner2012}%
  \BibitemOpen
  \bibfield  {author} {\bibinfo {author} {\bibfnamefont {S.}~\bibnamefont
  {Wagner}}, \bibinfo {author} {\bibfnamefont {O.}~\bibnamefont {Shishkina}}, \
  and\ \bibinfo {author} {\bibfnamefont {C.}~\bibnamefont {Wagner}},\
  }\href@noop {} {\bibfield  {journal} {\bibinfo  {journal} {J. Fluid Mech.}\
  }\textbf {\bibinfo {volume} {697}},\ \bibinfo {pages} {336} (\bibinfo {year}
  {2012})}\BibitemShut {NoStop}%
\bibitem [{\citenamefont {Bailon-Cuba}\ \emph {et~al.}(2012)\citenamefont
  {Bailon-Cuba}, \citenamefont {Shishkina}, \citenamefont {Wagner},\ and\
  \citenamefont {Schumacher}}]{Bailon2012}%
  \BibitemOpen
  \bibfield  {author} {\bibinfo {author} {\bibfnamefont {J.}~\bibnamefont
  {Bailon-Cuba}}, \bibinfo {author} {\bibfnamefont {O.}~\bibnamefont
  {Shishkina}}, \bibinfo {author} {\bibfnamefont {C.}~\bibnamefont {Wagner}}, \
  and\ \bibinfo {author} {\bibfnamefont {J.}~\bibnamefont {Schumacher}},\
  }\href@noop {} {\bibfield  {journal} {\bibinfo  {journal} {Phys. Fluids}\
  }\textbf {\bibinfo {volume} {24}},\ \bibinfo {pages} {107101} (\bibinfo
  {year} {2012})}\BibitemShut {NoStop}%
\bibitem [{\citenamefont {Shishkina}\ and\ \citenamefont
  {Wagner}(2012)}]{Shishkina2012}%
  \BibitemOpen
  \bibfield  {author} {\bibinfo {author} {\bibfnamefont {O.}~\bibnamefont
  {Shishkina}}\ and\ \bibinfo {author} {\bibfnamefont {C.}~\bibnamefont
  {Wagner}},\ }\href@noop {} {\bibfield  {journal} {\bibinfo  {journal} {J.
  Turbulence}\ }\textbf {\bibinfo {volume} {13}},\ \bibinfo {pages} {1}
  (\bibinfo {year} {2012})}\BibitemShut {NoStop}%
\bibitem [{\citenamefont {Shraiman}\ and\ \citenamefont
  {Siggia}(1990)}]{Shraiman1990}%
  \BibitemOpen
  \bibfield  {author} {\bibinfo {author} {\bibfnamefont {B.~I.}\ \bibnamefont
  {Shraiman}}\ and\ \bibinfo {author} {\bibfnamefont {E.~D.}\ \bibnamefont
  {Siggia}},\ }\href@noop {} {\bibfield  {journal} {\bibinfo  {journal} {Phys.
  Rev. A}\ }\textbf {\bibinfo {volume} {42}},\ \bibinfo {pages} {3650}
  (\bibinfo {year} {1990})}\BibitemShut {NoStop}%
\bibitem [{\citenamefont {Ching}(1997)}]{Ching1997}%
  \BibitemOpen
  \bibfield  {author} {\bibinfo {author} {\bibfnamefont {E.~S.~C.}\
  \bibnamefont {Ching}},\ }\href@noop {} {\bibfield  {journal} {\bibinfo
  {journal} {Phys. Rev. E}\ }\textbf {\bibinfo {volume} {55}},\ \bibinfo
  {pages} {1189} (\bibinfo {year} {1997})}\BibitemShut {NoStop}%
\bibitem [{\citenamefont {Grossmann}\ and\ \citenamefont
  {Lohse}(2012)}]{Grossmann2012}%
  \BibitemOpen
  \bibfield  {author} {\bibinfo {author} {\bibfnamefont {S.}~\bibnamefont
  {Grossmann}}\ and\ \bibinfo {author} {\bibfnamefont {D.}~\bibnamefont
  {Lohse}},\ }\href@noop {} {\bibfield  {journal} {\bibinfo  {journal} {Phys.
  Fluids}\ }\textbf {\bibinfo {volume} {24}},\ \bibinfo {pages} {125103}
  (\bibinfo {year} {2012})}\BibitemShut {NoStop}%
\bibitem [{\citenamefont {{Ahlers}}\ \emph {et~al.}(2014)\citenamefont
  {{Ahlers}}, \citenamefont {{Bodenschatz}},\ and\ \citenamefont
  {{He}}}]{Ahlers2014}%
  \BibitemOpen
  \bibfield  {author} {\bibinfo {author} {\bibfnamefont {G.}~\bibnamefont
  {{Ahlers}}}, \bibinfo {author} {\bibfnamefont {E.}~\bibnamefont
  {{Bodenschatz}}}, \ and\ \bibinfo {author} {\bibfnamefont {X.}~\bibnamefont
  {{He}}},\ }\href@noop {} {\bibfield  {journal} {\bibinfo  {journal} {J. Fluid
  Mech.}\ }\textbf {\bibinfo {volume} {758}},\ \bibinfo {pages} {436} (\bibinfo
  {year} {2014})}\BibitemShut {NoStop}%
\bibitem [{\citenamefont {{O. Shishkina, C. Wagner, C. R. Mecanique {\bf 333},
  17 (2005); O. Shishkina, C. Wagner, Comp. Fluids {\bf 36}, 484 (2007); S.
  Horn et. al., J. Fluid Mech. {\bf 724}, 175 (2013)}}()}]{flowsi2}%
  \BibitemOpen
  \bibfield  {author} {\bibinfo {author} {\bibnamefont {{O. Shishkina, C.
  Wagner, C. R. Mecanique {\bf 333}, 17 (2005); O. Shishkina, C. Wagner, Comp.
  Fluids {\bf 36}, 484 (2007); S. Horn et. al., J. Fluid Mech. {\bf 724}, 175
  (2013)}}},\ }\href@noop {} {\ }\BibitemShut {NoStop}%
\bibitem [{\citenamefont {{U. Schumann, J. Comput. Phys. {\bf 18}, 376 (1975);
  L. Schmitt, R. Friedrich, Notes Num. Fluid Mech. {\bf 20}, 355 (1988); C.
  Wagner et. al., Phys. Fluids {\bf 6}, 1425 (1994)}}()}]{flowsi1}%
  \BibitemOpen
  \bibfield  {author} {\bibinfo {author} {\bibnamefont {{U. Schumann, J.
  Comput. Phys. {\bf 18}, 376 (1975); L. Schmitt, R. Friedrich, Notes Num.
  Fluid Mech. {\bf 20}, 355 (1988); C. Wagner et. al., Phys. Fluids {\bf 6},
  1425 (1994)}}},\ }\href@noop {} {\ }\BibitemShut {NoStop}%
\bibitem [{\citenamefont {Shishkina}\ \emph {et~al.}(2010)\citenamefont
  {Shishkina}, \citenamefont {Stevens}, \citenamefont {Grossmann},\ and\
  \citenamefont {Lohse}}]{Shishkina2010}%
  \BibitemOpen
  \bibfield  {author} {\bibinfo {author} {\bibfnamefont {O.}~\bibnamefont
  {Shishkina}}, \bibinfo {author} {\bibfnamefont {R.~J. A.~M.}\ \bibnamefont
  {Stevens}}, \bibinfo {author} {\bibfnamefont {S.}~\bibnamefont {Grossmann}},
  \ and\ \bibinfo {author} {\bibfnamefont {D.}~\bibnamefont {Lohse}},\
  }\href@noop {} {\bibfield  {journal} {\bibinfo  {journal} {New J. Phys.}\
  }\textbf {\bibinfo {volume} {12}},\ \bibinfo {pages} {075022} (\bibinfo
  {year} {2010})}\BibitemShut {NoStop}%
\end{thebibliography}%
\end{document}